\documentclass[10pt]{IEEEtran}
\usepackage{cite}
\usepackage{placeins}
\usepackage{amsmath,amssymb,amsfonts}
\pdfoutput = 1
\usepackage{graphicx}
\usepackage{textcomp}
\usepackage{lipsum, color}
\usepackage{graphics}
\newcommand{\subparagraph}{}
\usepackage{titlesec}
\usepackage{subfig}
\usepackage{graphicx}
\usepackage{xcolor}
\usepackage{titlesec}
\usepackage{subfig}
\usepackage{graphicx}
\usepackage{mathtools}
\usepackage{soul}
\usepackage{balance}
\usepackage{algpseudocode}
\usepackage{float}
\usepackage{cite}
\usepackage[font=footnotesize]{caption}
\usepackage{tabularx}
\usepackage{array}
\usepackage[english]{babel}
\usepackage[utf8x]{inputenc}
\usepackage{amsmath}
\usepackage{multirow}
\usepackage{xcolor}
\usepackage{tikz}
\usepackage{blindtext}
\usepackage{enumitem}
\IEEEoverridecommandlockouts

\titlespacing\section{0pt}{4pt}{2pt}
\titlespacing\subsection{0pt}{3pt}{1pt}
\titlespacing\subsubsection{0pt}{3pt}{1pt}
\def\BibTeX{{\rm B\kern-.05em{\sc i\kern-.025em b}\kern-.08em
    T\kern-.1667em\lower.7ex\hbox{E}\kern-.125emX}}

\begin{document}
\title{MF-Net: Compute-In-Memory SRAM for Multibit Precision Inference using Memory-immersed Data Conversion and Multiplication-free Operators}

\author{Shamma Nasrin, ~\IEEEmembership{Student Member,~IEEE,} Diaa Badawi, Ahmet Enis Cetin,~\IEEEmembership{Fellow,~IEEE,} Wilfred Gomes, and Amit Ranjan Trivedi, ~\IEEEmembership{Member,~IEEE,}} 

\maketitle

\begin{abstract}
We propose a co-design approach for \textit{compute-in-memory} inference for deep neural networks (DNN). We use multiplication-free function approximators based on $\ell_1$ norm along with a co-adapted processing array and compute flow. Using the approach, we overcame many deficiencies in the current \emph{art} of in-SRAM DNN processing such as the need for digital-to-analog converters (DACs) at each operating SRAM row/column, the need for high precision analog-to-digital converters (ADCs), limited support for multi-bit precision weights, and limited vector-scale parallelism. Our co-adapted implementation seamlessly extends to multi-bit precision weights, it doesn't require DACs, and it easily extends to higher vector-scale parallelism. We also propose an SRAM-immersed successive approximation ADC (SA-ADC), where we exploit the parasitic capacitance of bit lines of SRAM array as a capacitive DAC. Since the dominant area overhead in SA-ADC comes due to its capacitive DAC, by exploiting the intrinsic parasitic of SRAM array, our approach allows low area implementation of within-SRAM SA-ADC. Our 8$\times$62 SRAM macro, which requires a 5-bit ADC, achieves $\sim$105 tera operations per second per Watt (TOPS/W) with 8-bit input/weight processing at 45 nm CMOS. Our 8$\times$30 SRAM macro, which requires a 4-bit ADC, achieves $\sim$84 TOPS/W. SRAM macros that require lower ADC precision are more tolerant of process variability, however, have lower TOPS/W as well. We evaluated the accuracy and performance of our proposed network for MNIST, CIFAR10, and CIFAR100 datasets. We chose a network configuration which adaptively mixes multiplication-free and regular operators. The network configurations utilize the multiplication-free operator for more than 85\% operations from the total. The selected configurations are 98.6\% accurate for MNIST, 90.2\% for CIFAR10, and 66.9\% for CIFAR100. Since most of the operations in the considered configurations are based on proposed SRAM macros, our compute-in-memory's efficiency benefits broadly translate to the system-level.    
\end{abstract}
\begin{IEEEkeywords}
Deep neural networks; In-SRAM processing; MNIST; CIFAR10; CIFAR100.
\end{IEEEkeywords}
\footnote{Shamma Nasrin (snasri2@uic.edu), Diaa Badawi(dbadaw2@uic.edu), Ahmet Enis Cetin(aecyy@uic.edu) and Amit Trivedi(amitrt@uic.edu) are with the department of Electrical and Computer Engineering, University of Illinois at Chicago. Wilfred Gomes(wilfred.gomes@intel.com) is affiliated with Intel.  }

\section{Introduction}
In many practical applications, deep neural networks (DNNs) have shown a remarkable prediction accuracy \cite{makantasis2015deep,liu2015targeting,wei2017boosting,cho2015much}. DNNs in these applications typically utilize thousands to millions of parameters (i.e., weights) and are trained over a huge number of example patterns \cite{qassim2017residual,carvalho2017exposing,szegedy2017inception}. Operating over such a large parametric space, which is carefully orchestrated over multiple abstraction levels (i.e., hidden layers), facilitates DNNs with a superior generalization and learning capacity, but also presents critical inference constraints, especially when considering real-time and/or low power applications. For instance, when DNNs are mapped on a traditional computing engine, the inference performance is strangled by extensive memory accesses, and the high performance of the processing engine helps little.
\begin{figure}[t]
	\centering
	\includegraphics[width=\linewidth]{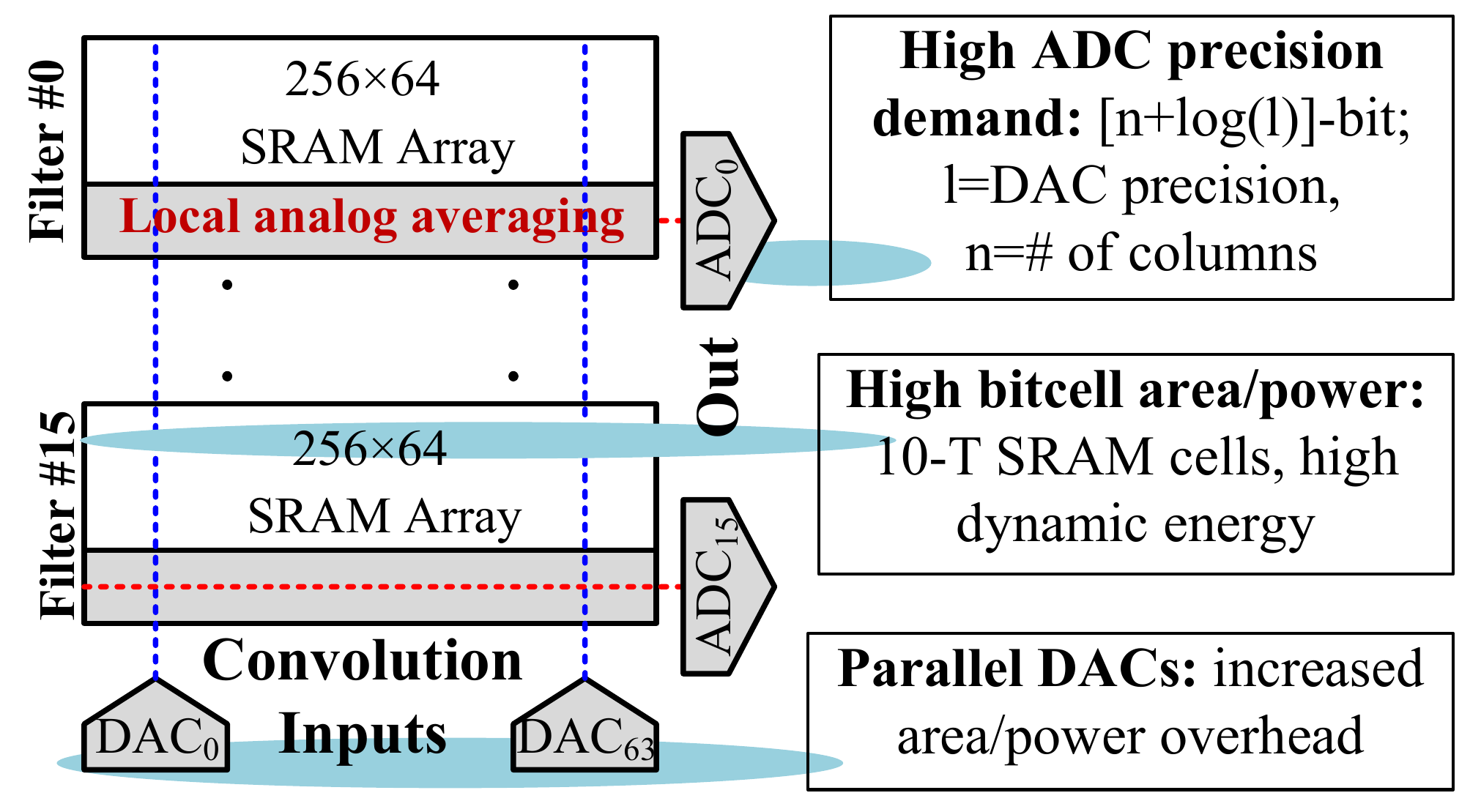}
	\caption{High-level overview of in-SRAM processing in the current art and key limitations. The figure considers CONV-SRAM \cite{biswas2018conv} as a motivating example, however, the challenges are common to the most other designs.}
	\label{fig:1}
\end{figure}

A radical approach, gaining attention to address this performance challenge of DNN, is to design memory units that can not only store DNN weights but also locally process DNN layers. Therefore, using such `compute-in-memory' high volume data traffic between processor and memory units is obviated, and the critical bottleneck can be alleviated. Moreover, a \emph{mixed-signal} in-memory processing of DNN operands reduces necessary operations for DNN inference. For example, using charge/current-based representation of the operands, the accumulation of products simply reduces to current/charge summation over a wire. Therefore, dedicated modules and operation cycles for product summations are not necessary. 

In recent years, several in-SRAM DNN implementations have been shown. However, many critical limitations remain, which inhibit the scalability of the processing. In Figure 1, using CONV-SRAM \cite{biswas2018conv} as a motivating example, we highlight these limitations -- notably, the challenges are common to most other in-SRAM applications too. To compute the inner product of $l$-element weight and input vectors $\mathbf{w}$ and $\mathbf{x}$, $l$-DACs and one ADC are required. Since DACs are concurrently active, they lead to both high area and power. With the increasing precision of operands, the design of DACs also becomes more complex. In \cite{kang2018memory}, time-domain DACs are used to handle this complexity; however, with increasing input precision, either operating time increases exponentially, or complex analog-domain voltage scaling is necessitated. In \cite{jiang2019xnor}, DACs are obviated, but the operation is only limited to binary inputs and weights, which has low accuracy.

An ADC is needed to digitize the inner product of $\mathbf{w}$ and $\mathbf{x}$ vectors in Figure 1. If $\mathbf{x}$ is $n$-bit and ADC combines the output of $l$ cells, the minimum necessary precision of the ADC is $\sim n+log_2(l)$ to avoid any quantization loss. Therefore, ADC precision requirement becomes more stringent with increasing input precision and the number of cells being summed. Moreover, scaled technology nodes of SRAM precludes analog-heavy ADCs embedded within SRAM. In \cite{biswas2018conv}, a charge sharing-based ADC was integrated with SRAM. However, the worst-case comparison steps grow exponentially with ADC's precision, limiting vector scale parallelism (i.e., the number of cells/products $l$ that can be concurrently processed). In \cite{zhang2017memory}, ADC is avoided by using a comparator circuit, but this limits the implementation only to step function-based activation and doesn't support the mapping of DNNs with larger weight matrices that cannot fit within an SRAM array.

 \begin{figure*}[t]
	\centering
	\subfloat[]{\includegraphics[width=0.33\linewidth]{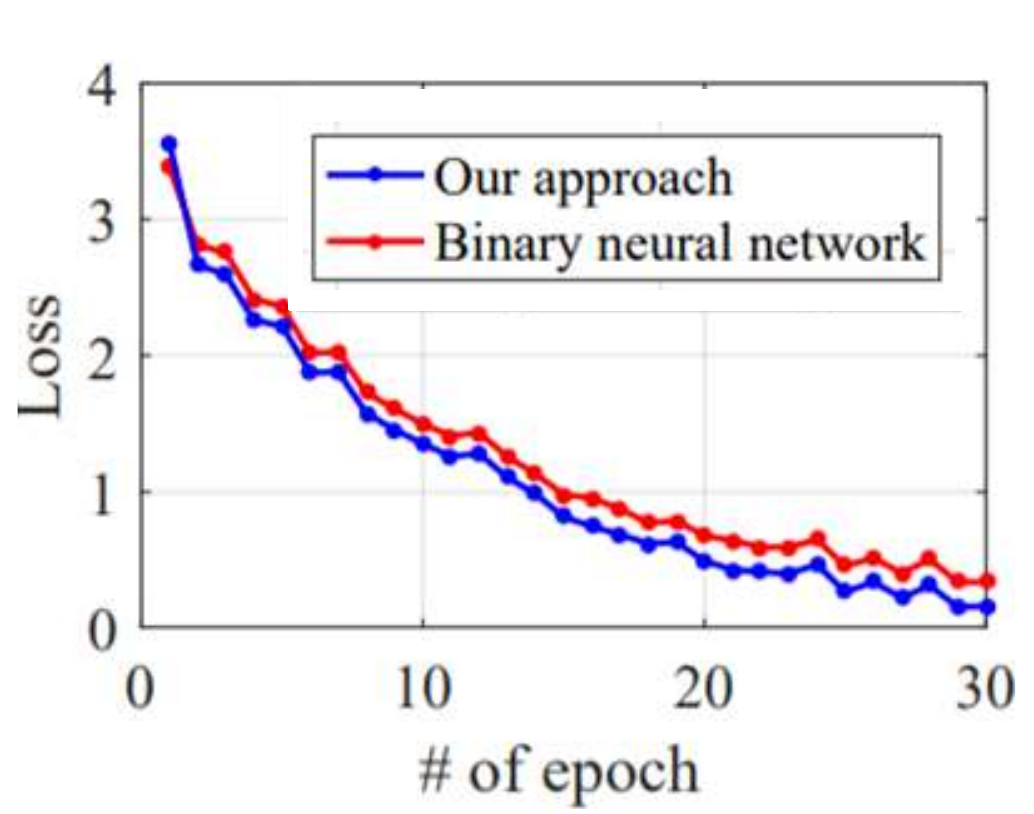}}
	\subfloat[]{\includegraphics[width=0.33\linewidth]{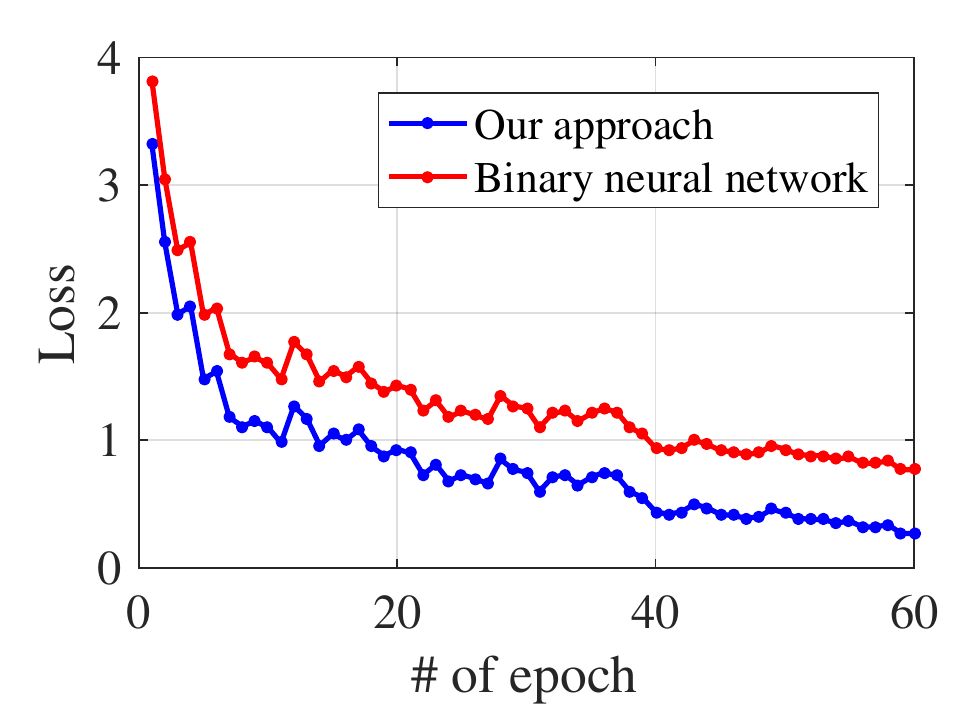}}
	\subfloat[]{\includegraphics[width=0.33\linewidth]{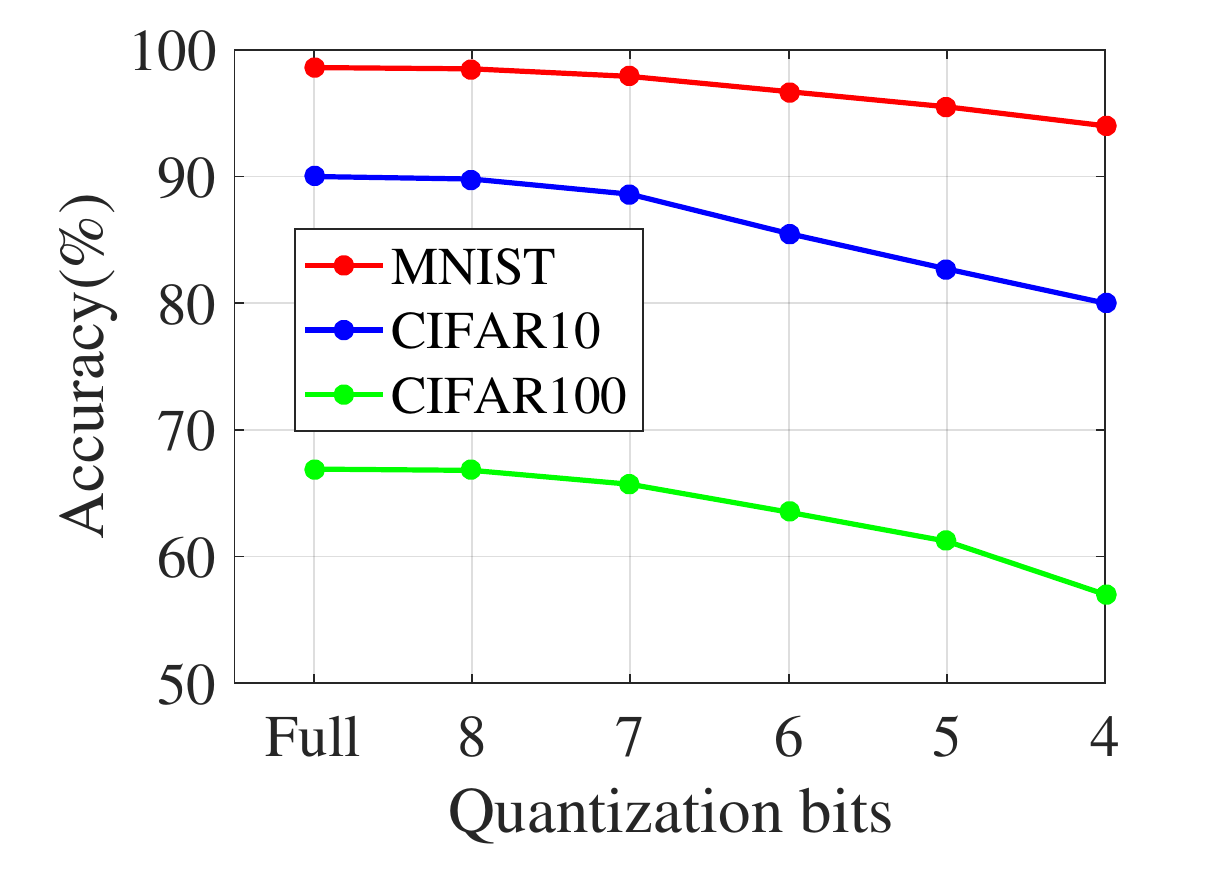}}
	\caption{Training curves for the multiplication-free NN operator on (a) MNIST and (b) CIFAR10 data sets. (c) The effect of quantization on multiplication-free operator.}
	\label{fig:acc_results}
\end{figure*}
Near-memory processing avoids the complexity of ADC/DAC by operating in the digital domain only. The schemes use the time-domain and frequency-domain summing of weight-input products. Unlike charge/current-based sum, however, time/frequency-domain summation is not instantaneous. A counter \cite{amaravati201855} or memory delay line (MDL) \cite{sayal201912} can be used to accumulate weight-input products. With increasing vector-scale parallelism (length of input/weight vector $l$), the integration time of counter/MDL increases exponentially, which again limits parallelism and throughput.

Addressing the challenges of state-of-the-art, we propose the following core concepts:

\begin{enumerate}[leftmargin=4mm,itemsep=0pt,topsep=0pt,parsep=0pt,partopsep=0pt]
\item We use a multiplication-free neural network operator that eliminates high-precision multiplications in input-weight correlation\cite{pan,ergen,afrasiyabi}. In the new operator, the correlation of weight $\mathbf{w}$ and input $\mathbf{x}$ is represented as
\begin{equation} \label{eq:1}
    \mathbf{w} \oplus \mathbf{x} = \sum_i \text{sign}(x_i)\cdot \text{abs}(w_i) + \text{sign}(w_i)\cdot \text{abs}(x_i)
\end{equation}
Here, $\cdot$ is an element-wise multiplication operator, $+$ is element-wise addition operator, and $\sum$ is vector sum operator. $\text{sign}()$ operator is $\pm 1$ and $\text{abs}()$ operator produces absolute unsigned value of the operand. Therefore, in Eq. \ref{eq:1}, the correlation operator is \emph{inherently} designed to only multiply a one-bit element of $\text{sign}(\mathbf{x})$ against full precision $\mathbf{w}$, and one-bit $\text{sign}(\mathbf{w})$ against $\mathbf{x}$. By avoiding direct multiplications between full precision variables, DACs can be avoided in in-memory computing. 

\item Additionally, we reformulate Eq. \ref{eq:1} as below to minimize the dynamic energy of computation:  
\begin{subequations} \label{Eq:2}
\begin{equation}
\begin{aligned}
\sum_i \text{sign}(w_i)\cdot \text{abs}(x_i) = \\ 2\times \underbrace{\sum_i \text{step}(w_i)\cdot \text{abs}(x_i)}_\text{low dynamic energy} - \underbrace{\sum_i \text{abs}(x_i)}_\text{shared computation}  
\end{aligned}
\end{equation}
\begin{equation}
\begin{aligned}
\sum_i \text{sign}(x_i)\cdot\text{abs}(w_i) = \\ 2\times \underbrace{\sum_i \text{step}(x_i)\cdot\text{abs}(w_i)}_\text{low dynamic energy} - \underbrace{\sum_i \text{abs}(w_i)}_\text{weight statistics}  
\end{aligned}
\end{equation}
\end{subequations}
In the above reformulation, $\text{step}() \in [0, 1]$. The above reformulation allows processing with single product port of SRAM cells; thus, it can reduce dynamic energy. Compare this to current implementations \cite{zhang2017memory, biswas2018conv, kang2018memory} where operations with weights $w \in [-1, 1]$ require product accumulation over both bit lines. SRAM design in \cite{biswas2018conv} is 10T to support differential ended processing, while our SRAM is 8T due to single-ended processing. However, the above reformulation also has residue terms $\sum \text{abs}(x_i)$ and $\sum \text{abs}(w_i)$. The first term can be computed using a dummy row of weights, all storing ones. For a given input, this computation is referenced for all weight vectors; thus, computing overheads amortize. The second term is a weight statistics that can be pre-computed and can be looked-up during evaluation. 

\item We also discuss that parasitic capacitance of bit lines of SRAM array can be exploited as a capacitive digital-to-analog converter (DAC) for successive approximation-based ADC (SA-ADC). In our architecture, when bit lines in one half of the array compute the weight-input correlation, bit lines in the other half implement binary search of SA-ADC to digitize the correlation output. Remarkably, the proposed DNN operator also helps reducing precision constraints on SA-ADC. With the proposed operator, each SRAM cell only performs 1-bit logic operation; thus, to digitize the output of $l$ columns, ADC with $log_2(l)$ precision is needed. Compare this to CONV-SRAM in Figure 1, where necessary ADC's precision is $n+log_2(l)$ since each SRAM cell processes $n$-bit DAC's output. By simplifying data converters, our scheme can also achieve higher vector-scale parallelism, i.e., allows processing a higher number of parallel columns ($l$) with the same ADC complexity than in \cite{biswas2018conv}.

\end{enumerate}

Sec. II introduces the co-adapted multiplication-free operators for the in-SRAM deep neural network. Sec. III characterizes the algorithmic accuracy of our framework. Sec. IV gives an overview of compute-in-memory macro based on multiplication-free operator. Sec. V discusses power/performance characterization of the proposed compute-in-SRAM macro. Sec. VI explores the concept of synergistic integration of digital and compute-in-memory processing for DNN. Sec. VII concludes. 

\section{Co-designed Multiplication-free NN Operator}
Muliplication-free DNN operator were presented in \cite{akbacs2015multiplication}. In this work, we expand on the potential of multiplication-free operators to considerably reduce the complexity of SRAM-based compute-in-memory design. Compared to \cite{akbacs2015multiplication}, we adjusted the operator with abs() operation on operands $w$ and $x$ in Eq. (1) to further simplify compute-in-memory processing steps. Our later discussion will show that the adjusted operator also achieves high prediction accuracy on various benchmark datasets. Note that a multiplication-free operator in Eq. (1) is based on the $\ell_1$ norm, since $\mathbf{x} \oplus \mathbf{x} = 2 ||x||_1$. In traditional neural networks, neurons perform inner products to compute the correlation between the input vector with the weights of the neuron. We define a new neuron by replacing the affine transform of a traditional neuron using co-designed NN operator as $\phi(\mathbf{\alpha} (\mathbf{x} \oplus \mathbf{w}) + b)$ where $\mathbf{w}\in {R}^{d}$, $\alpha, b\in {R}$ are weights, the scaling coefficient and the bias, respectively. Moreover, since the proposed NN operator is nonlinear itself, an additional nonlinear activation layer (e.g., ReLU) is not needed, i.e., $\phi()$ can be an identity function. Most neural network structures including multi-layer perceptrons (MLP), recurrent neural networks (RNN), and convolutional neural networks (CNN) can be easily converted into such an \emph{compute-in-memory compatible} network structures by just replacing ordinary neurons with the activation functions defined using $\oplus$ operations without modification of the topology and the general structure.

The above co-designed neural network can be trained using standard back-propagation and related optimization algorithms. The back-propagation algorithm computes derivatives with respect to the current values of parameters. However, the key training complexity for the operator is that the derivative of $\mathbf{\alpha} (\mathbf{x}  \oplus\mathbf{w}) + b$ with respect to $\mathbf{x}$ and $\mathbf{w}$ is undefined when $x_i$ and $w_i$ are zero. The partial derivative of $\mathbf{x} \oplus\mathbf{w}$ with respect to $x$ and $w$ can be expressed:
\begin{subequations} \label{Eq:3}
\begin{equation} 
	\frac{\partial (\mathbf{x}\oplus \mathbf{w})}{\partial x_i} =  \text{sign}(w_i)\text{sign}(x_i) + 2\times\text{abs}(w_i)\delta(x_i),
	\label{eq:op1}
\end{equation}
\begin{equation} 
	\frac{\partial (\mathbf{x}\oplus \mathbf{w})}{\partial w_i} =  \text{sign}(x_i)\text{sign}(w_i) + 2\times\text{abs}(x_i)\delta(w_i).
	\label{eq:op1}
\end{equation}
\end{subequations} 
Here, $\delta()$ is a Dirac-delta function. For gradient-descent steps, the discontinuity of sign function can be approximated by a steep hyperbolic tangent and the discontinuity of Dirac-delta function can be approximated by a steep zero-centered Gaussian function.   

\begin{table}[]
\centering
\caption{Multiplication-free operator vs. conventional DNN and  binary neural network (BNN))}
\label{tab:my-table}
\begin{tabular}{|l|c|c|c|}
\hline
\multirow{2}{*}{Dataset} & \multicolumn{3}{c|}{Accuracy on test set} \\ \cline{2-4} 
                         & Conventional & Multiplication-free & BNN  \\ \hline
MNIST                    & 99.01\%      & 98.6\%              & 97\% \\ \hline
CIFAR10                  & 90.95\%      & 90.2\%              & 85\% \\ \hline
CIFAR100                 & 67.88\%      & 66.9\%              & -    \\ \hline
\end{tabular}
\end{table}
\section{Accuracy on Benchmark Datasets}
In this section, we characterize the prediction accuracy of the proposed NN operator for MNIST \cite{mnist}, CIFAR10 \cite{cifar} and CIFAR100 \cite{cifar100} data-sets. For MNIST, we simulate the LeNet-5 network \cite{L5}. The network consists of two convolution layers, two fully connected layers, and uses max-pooling layers. For CIFAR10, we use a convolutional neural network consisting of five convolution layers and two fully connected layers. DNN for CIFAR10 uses max-pooling layers and batch normalization layer. For CIFAR100 we choose MobileNetV2 \cite{mobilenetv2}. Both MNIST and CIFAR-10 data sets consist of 60,000 images with 10 classes, whereas the CIFAR100 data set consists of 60,000 images with 100 classes. A test-set of 10,000 images is used to characterize the prediction accuracy. 

Table I summarizes test-set accuracy for the data sets with conventional, multiplication-free, and binarized network operators. In Table 1, for CIFAR10 results, the convolution layers are binarized in BNN and made multiplication-free in the proposed network, respectively, while the fully-connected layers are implemented using regular multiplications. Accuracy results for MNIST are obtained by keeping only the last layer typical, and the other layers are binarized or multiplication-free. For CIFAR100 results, we make only the bottleneck (BN) layers of MobileNetV2 multiplication-free and keep the rest of the network with the typical operator. In Table I, we also compare the accuracy of the multiplication-free operator against the binarized weight operator using the above configuration on various datasets. While a binary neural network (BNN) simplifies the implementation at the cost of constrained learning space, the learning space of the multiplication-free operator is as extensive as the typical operator. Therefore, multiplication-free operator-based networks have much better accuracy than BNN. The accuracy of the multiplication-free operator is also quite competitive to the conventional operator on various datasets\cite{pan,ergen}. Meanwhile, our later discussion will show that a multiplication-free operator significantly reduces implementation complexity compared to the typical operator. Figures \ref{fig:acc_results}(a-b) show the training curve for the multiplication-free operator compared to BNN under identical training conditions for MNIST and CIFAR10. In Figure \ref{fig:acc_results}(c), the prediction accuracy of multiplication-free operator is also amenable to input/weight quantization. Operation with 8-bit fixed precision inputs/weights has equivalent accuracy to the floating-point precision cases on various datasets. Therefore, in the following discussion, we consider an 8-bit fixed-precision operation with the multiplication-free operator. 

\begin{figure*}[t]
	\centering
	\includegraphics[width=1\linewidth]{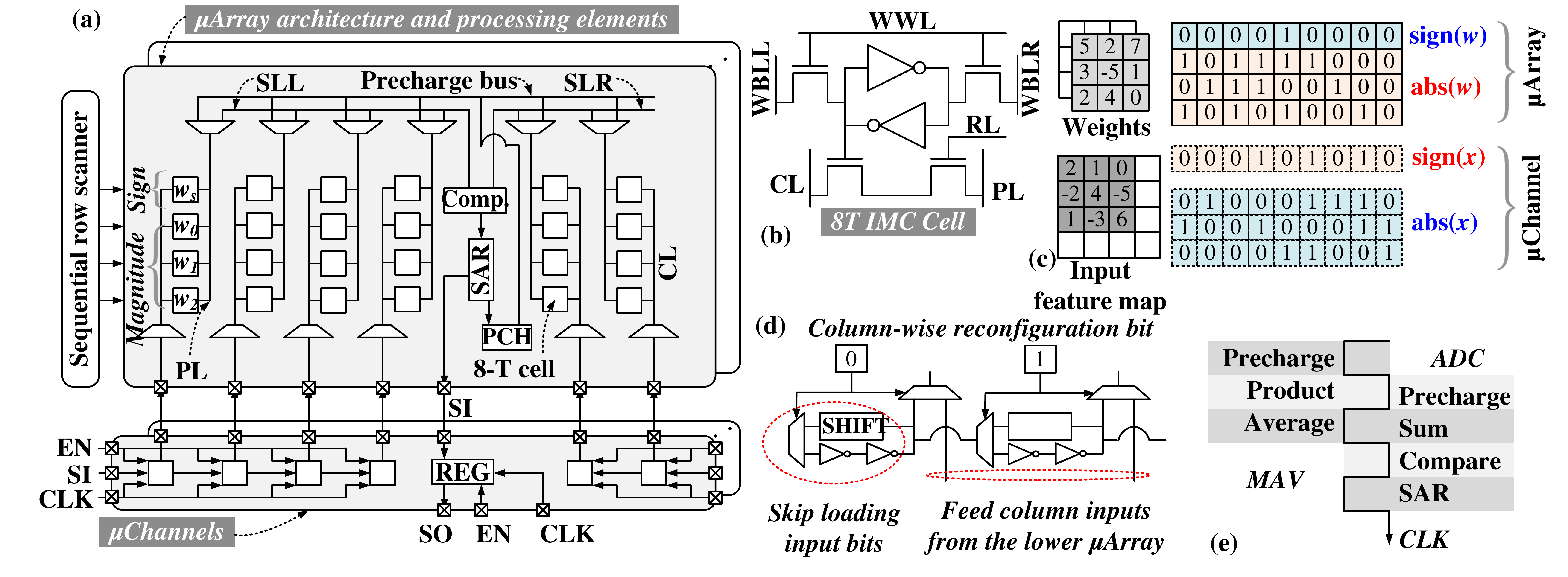}\\
	\caption{(a) $\mu$Array and $\mu$Channel architecture for $\mathbf{w}\oplus\mathbf{x}$ processing in a compute-in-SRAM macro. (b) 8-T SRAM cell for in-memory processing. SRAM cell has additional transistors, colored red, for input-weight correlation. (c) Mapping of sign bits and the absolute values of weights and inputs on $\mu$Arrays and $\mu$Channels. (d) Stitching of $\mu$Array columns by reconfiguration bits in $\mu$Channels. (e) Instruction cycles for in-SRAM processing.}
	\label{fig:complete}
\end{figure*}

\section{Overview of Compute-in-SRAM Macro based on Multiplication-free Operator}

\subsection{Compute-in-SRAM macro based on $\mu$Arrays/Channels}
Figure \ref{fig:complete}(a) shows the proposed design of compute-in-SRAM macro for multiplication-free operator-based DNN inference. In the proposed design, an SRAM macro consists of $\mu$Arrays and $\mu$Channels, as shown in the figure. Each $\mu$Array is dedicated to storing one weight channel. DNN weights are arranged across columns in a $\mu$Array where each bit plane of weights is arranged in a row. Therefore, an $N$-dimensional weight channel with $m$-bit precision weights will require $m$ rows and $N$ columns of SRAM cells in a $\mu$Array. Figure \ref{fig:complete}(b) shows the proposed 8T SRAM cell used for the in-SRAM processing of the operator. Extra transistors in the cell compared to a 6T cell decouple typical read/write operations to within cell product. The added transistors are selected by the row and column select lines (RL and CL) and operate on the product bit line (PL). The decoupling of read/write and product operations mitigates interference between the operations, reduces the impact of process variability, and allows operation in storage hold mode. Each $\mu$Array is augmented with a $\mu$Channel. $\mu$Channels convey digital inputs/outputs to/from $\mu$Arrays. $\mu$Channels are essentially low overhead serial-in serial-out digital paths based on scan-registers. If a weight filter has many channels, $\mu$Channels also allow stitching of $\mu$Arrays so that inputs can be shared among the $\mu$Arrays. Figure \ref{fig:complete}(d) shows $\mu$Channel-based column merging between two $\mu$Arrays. If two columns are merged, inputs are passed to the top array directly from the bottom array, and the loading of input bits is bypassed on the top column; therefore, overheads to load input feature-map are minimized. Figure \ref{fig:complete}(c) illustrates input/weight mapping to SRAM macro and operation sequence. For $\text{step}(\mathbf{x})\cdot \text{abs}(\mathbf{w})$ step in $\mathbf{w}\oplus\mathbf{x}$, $\text{step}(\mathbf{x})$ vector is loaded on the $\mu$Channel and operated against $\text{abs}(\mathbf{w})$ rows of $\mu$Array. For $\text{step}(\mathbf{w})\cdot \text{abs}(\mathbf{x})$ step, bitplanes of $\text{abs}(\mathbf{x})$ vector are sequentially loaded on the $\mu$Channel and operated against $\text{step}(\mathbf{w})$ row of the $\mu$Array.   

\subsection{Operation cycles}
\begin{figure*}[t]
	\centering
	\includegraphics[width=\linewidth]{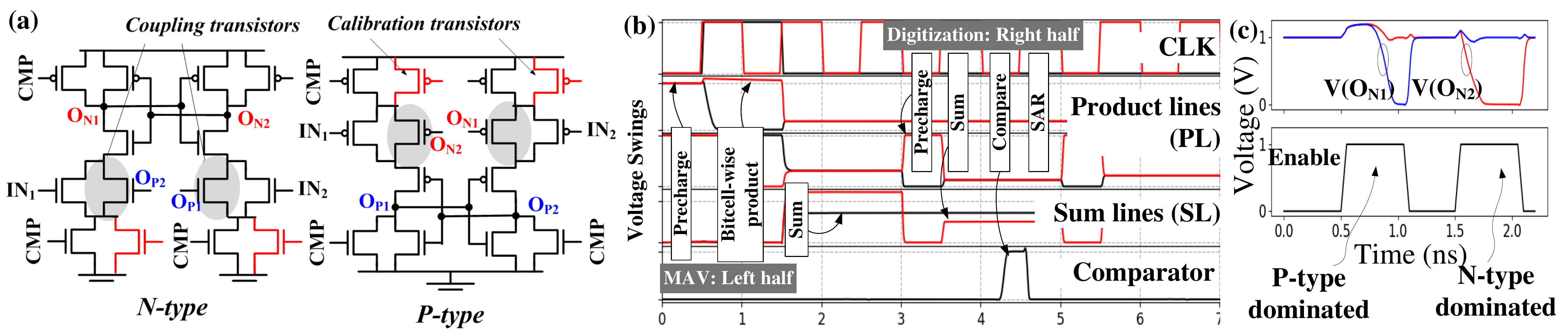}
	\caption{(a) Cross-coupled comparator schematic. Operational transients for (b) in-SRAM processing of multiplication-free operator and (c) comparator.}
	\label{fig:transients1}
\end{figure*}

In a $\mu$Array, to compute $\mathbf{x} \oplus \mathbf{w}$, the operation proceeds by bit planes. If the left half computes the weight-input product, the right half digitizes. Both halves subsequently exchange their operating mode to process weights stored in the right half. When evaluating the inner product terms $\text{step}(\mathbf{x})\cdot \text{abs}(\mathbf{w})$, computations for i$^{th}$ weight vector bitplane are performed in one instruction cycle. At the start, the inverted logic values of $\text{step}(\mathbf{x})$ bit vector is applied to CL through $\mu$Channels. PL is precharged. When clock switches, tri-state MUXes float PL. Compute-in-memory controller activates SRAM rows storing i$^{th}$ bit vector of $\mathbf{w}$. In a column $j$, only if both $w_{j,i}$ and $\text{step}(x_{j})$ are one, the corresponding PL segment discharges. To minimize the leakage power, we maintain SRAM cells in their hold mode and dedicate additional clock time to discharge PLs. The potential of all column lines is averaged on the sum-lines to determine the net multiply-average (MAV), i.e., $\sum \frac{1}{N}\big(w_{j,i}\times \text{step}(x_{j})\big)$ for input vector and weight bit plane $\mathbf{w_j}$. Figure \ref{fig:complete}(e) shows the instruction sequence for the left half to compute MAV consisting of precharge, product, and average stages. Figure \ref{fig:transients1}(b) shows the simulated transients in the left half to illustrate the execution sequence. $\text{step}(\mathbf{w})\cdot \text{abs}(\mathbf{x})$ is processed similarly by loading bitplanes of $\text{abs}(\mathbf{x})$ to the $\mu$Channels.  

\subsection{SRAM-Immersed SA-ADC}
\begin{figure}[t]
	\centering
    \includegraphics[width=0.8\linewidth]{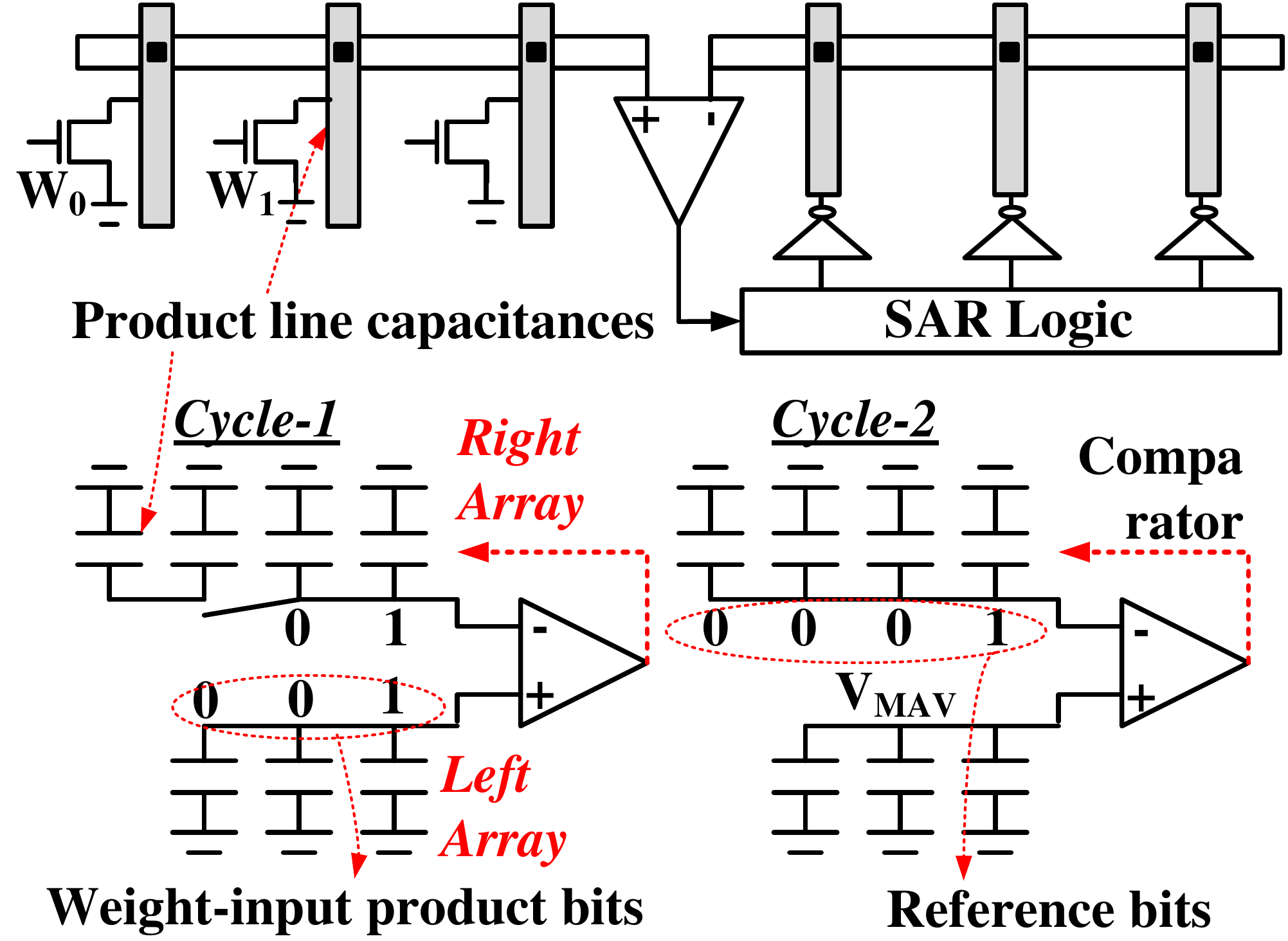}
	\caption{SRAM-Immersed successive approximation (SA) ADC exploits bitline parasitic as capacitive digital-to-analog converter (DAC). }
	\label{fig:ADC}
\end{figure}

Since MAV output at the sum line ($SL$) is charge-based, an analog-to-digital converter (ADC) is necessary to convert the output into digital bits. In Figure \ref{fig:complete}(a), the right half of the array implements an SRAM-immersed successive approximation (SA) data converter to digitize the output at the left sum line ($SLL$). Reference voltages for SA-based data conversion are generated by exploiting PL parasitic in the right half. Figure \ref{fig:ADC} describes the utilization of parasitic capacitance of the product lines for the DAC implementation of SA-ADC. The product lines of the right half are charged and discharged according to the SAR logic to produce the reference voltage at the right sum line (SLR). In the i$^{th}$ SA iteration, 2$^i$ capacitors are used to generate the reference voltage. Each half also uses a dummy PL of matching capacitance to complete SA. Although the capacitance of SL affects the MAV range, its effect nullifies during the digitization since the capacitor is a common mode to both ends of the comparator. Nonetheless, limited voltage swing range due to SL's capacitance limits the number of parallel columns in a $\mu$Array that can be reliably operated. 

Figure \ref{fig:complete}(e) also shows the instruction cycles for the data conversion consisting of precharge, average, compare, and SAR steps. One cycle of data conversion lasts two clock periods. For $n$-bit digitization, $2n$ clock cycles are needed. In a conversion cycle, at the start, PLs in the right half are charged based on initialization or SA output from the previous cycle. At the next clock transition, PLs are merged to average their voltage. Next, a comparator compares the potential at the left and right sum lines (SLL and SLR in Figure \ref{fig:complete}(a)). Subsequently, SA logic operates on the comparator's output to update the digitization registers and produces the next precharge logic bits. Figure \ref{fig:transients1}(b) shows the simulated transients in the right half and interplay with the left half. 

The comparator in our design must accommodate rail-to-rail input voltages at SLL and SLR. Therefore, in Figure \ref{fig:transients1}(a), we use a cross-coupled comparator consisting of n-type and p-type modules. The n-type module receives inputs at NMOS transistors while p-type receives at PMOS. Coupling transistors to integrate both modules are highlighted in Figure \ref{fig:transients1}(a). If the input voltages are closer to zero, the p-type instance dominates. Otherwise, if the input voltages are close to VDD, the n-type instance dominates. Connections to coupling transistors in the figure ensure that n-type or p-type instances can be overridden at the appropriate voltage range. Figure \ref{fig:transients1}(c) shows the comparator output transients. 

\subsection{Key advantages over the current compute-in-SRAM}
Our multiplication-free inference framework using compute-in-SRAM $\mu$Arrays and $\mu$Channels has many key advantages over the competitive designs \cite{1st,2nd,3rd}. \textit{First}, a multiplication-free learning operator, obviates digital-to-analog converters (DAC) in SRAM macros. Meanwhile, DACs incur considerable area/power in the current competitive designs \cite{conv,nasrin}. Although overheads of DAC can be amortized by operating in parallel over many channels, the emerging trends on neural architectures, such as depth-wise convolutions in MobileNets \cite{mobile}, show that these opportunities may diminish. Comparatively, our DAC-free framework is much more efficient in handling even \textit{thin} convolution layers by eliminating DACs; thereby, allowing fine-grained embedding of $\mu$Channels without considerable overheads. If the filter has many parallel channels, our architecture can also exploit input reuse opportunities by merging $\mu$Channels as we discussed earlier. \textit{Secondly}, a multiplication-free operator, is also synergistic with the discussed bitplane-wise processing. Bitplane-wise processing followed in this work reduces the ADC's precision demand in each cycle by limiting the dynamic range of MAV. Note that with bitplane-wise processing, for $n$ column lines, MAV varies over 2$^n$ levels. However, if such bitplane-wise processing is performed for the typical operator, an excessive $O(n^2)$ operating cycles will be needed for $n$-bit precision. Meanwhile, a multiplication-free operator only requires $O(2n)$ cycle. \textit{Lastly}, we have discussed unique opportunities to exploit SRAM array parasitic for SRAM-immersed ADC. The proposed scheme obviates a major area overhead for SA-ADC. Therefore, the compute-in-SRAM macro can maintain a high memory density.

\section{Power--Performance Characteristics of Compute-in-SRAM Macro}
\subsection{Simulation methodology}

In this section, we discuss power--performance characteristics of compute-in-SRAM macro presented earlier. We use LeNET-5 for MNIST characterization as a running use-case to discuss various design and power optimization opportunities. The scalability of the proposed framework on more complex datasets and deeper networks will be discussed in the next section. The network weights for LeNET-5 are trained using TensorFlow. Convolutional layers in LeNET-5 and the next fully-connected layer are implemented using a multiplication-free operator. The last layer of LeNET-5 is implemented in a typical operator. Data transfer among SRAM arrays and post-processing of array's output, such as applying max-pooling, is simulated functionally. Compute-in-SRAM operations are simulated using HSPICE in 45 nm CMOS technology using predictive technology models \cite{Zhao}. Each $\mu$Array in SRAM-macro has 8 rows and 62 columns. A $\mu$Array is split into halves, where each half stores a weighted channel by flattening it to one-dimensional. $\mu$Arrays process 8-bit weights against 8-bit inputs. In each SA cycle in $\mu$Arrays, MAV is digitized to 5-bit output. If the flattened filter width is more than 31, it is partitioned and mapped on more $\mu$Arrays. 

\begin{figure}[t]
\centering
\subfloat[]{\includegraphics[width=0.55\linewidth]{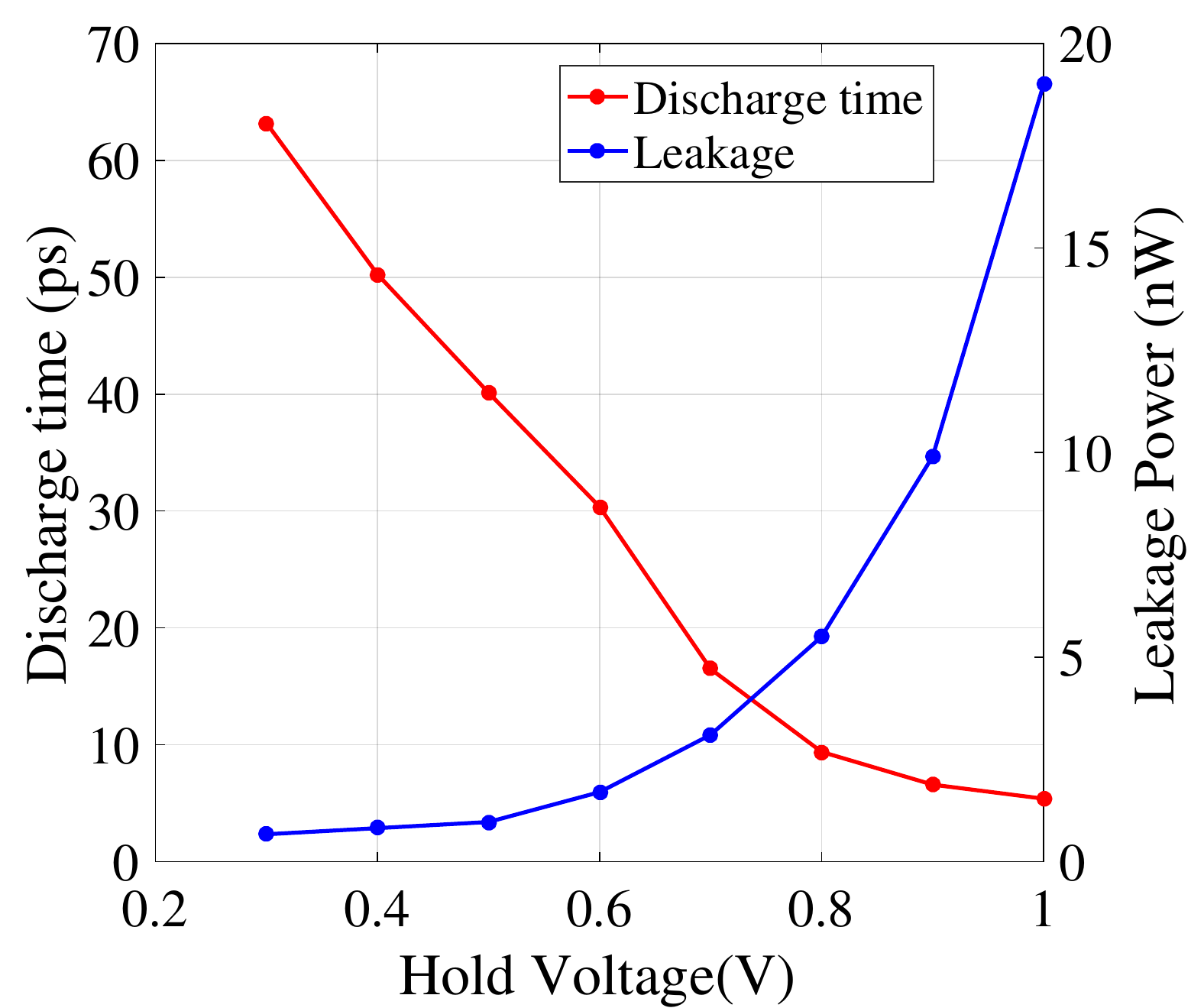}}
\subfloat[]{\includegraphics[width=0.45\linewidth]{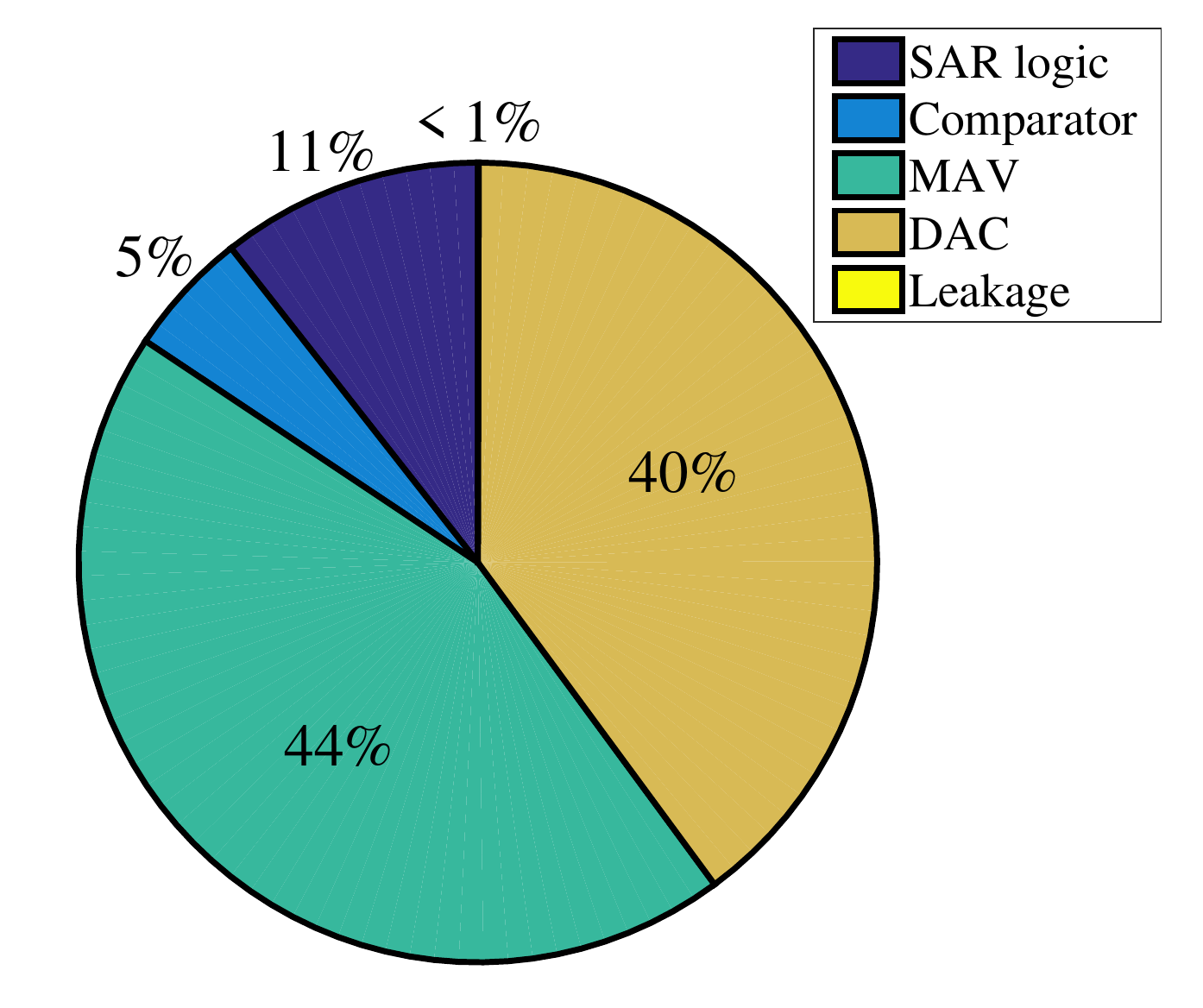}}

\caption{Power performance of SRAM array. (a) The product line discharge time and SRAM leakage power at varying hold voltage. (b) Distribution of dynamic and leakage energy for a $\mu$Array for performing MAV and within-SRAM digitization.}
\label{fig:leakage}
 \end{figure}

\subsection{Power--performance characterizations}
We summarize key power--performance characteristics of our compute-in-SRAM macros based on 45 nm CMOS simulations using PDK \cite{pdk}. In Figure \ref{fig:leakage}(a), reducing the hold voltage of SRAM cells reduces leakage current through the cells; however, also increases the PL discharge time during product computation. We chose the hold voltage to be 0.4 V since this represents an optimal balance of discharge time and leakage power. At this voltage, the worst case discharge time (considering slow-slow corner and 120$^0$C temperature) is 50 ps. The average leakage power of a $\mu$Array is 0.97 $n$W at the typical corner. Figure \ref{fig:leakage}(b) shows the distribution of power among various operations in a $\mu$Array. A $\mu$Array consumes $\sim$7.6 $\mu$W of active power to perform the MAV operation while operating at 1 GHz. MAV operation consumes 44\% of the total energy, whereas digitization consumes 55\% of the total energy, accounting for capacitive DAC, comparator, and SAR logic operations. The leakage power of SRAM accounts for $<$1\% of the total. Since ADC's power overhead are considerable, in the next section, we discuss techniques for dynamic precision scaling to mitigate the overhead. 

\begin{figure}[t]
	\centering
    \includegraphics[width=0.95\linewidth]{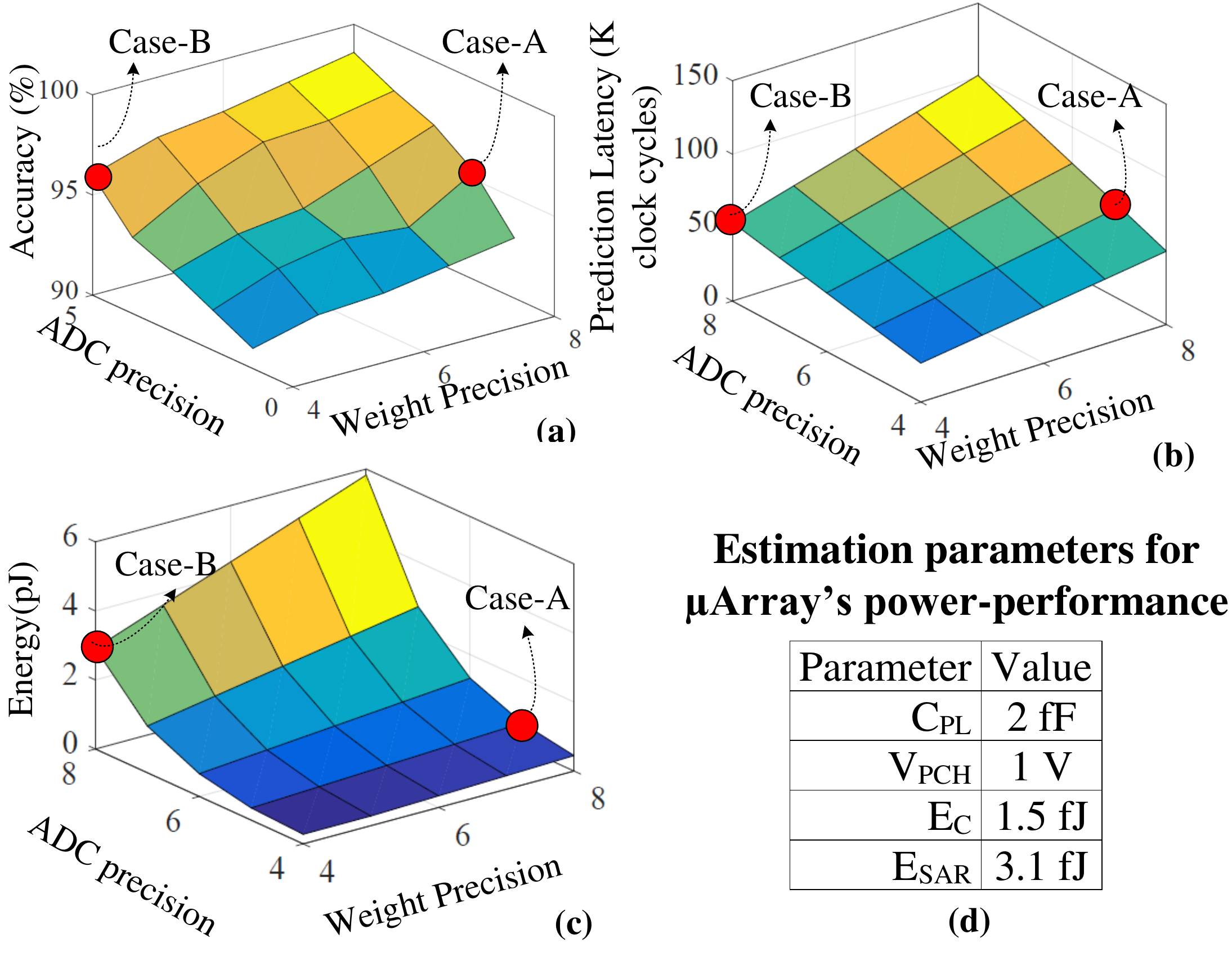}
	\caption{Surface plot for (a ) accuracy for MNIST characterization, and (b) latency and (c) energy of compute-in-SRAM macros at varying precision of weights and ADC. (d) Parameters used in Equation 4.  }
	\label{fig:surface}
\end{figure}

\subsection{Dynamic precision scaling }
In Figure \ref{fig:surface}, we explore the design space of varying weight precision ($W_P$) and ADC precision ($A_P$) for predictions on MNIST dataset. Figure \ref{fig:surface}(a) shows a surface plot showing the prediction accuracy at varying $W_P$ and $A_P$. The prediction accuracy improves with higher $W_P$ and $A_P$. Figures \ref{fig:surface}(b-c) show surface plot for latency and energy for a unit $\mathbf{w}\oplus\mathbf{x}$ operation on the $\mu$Array, respectively. The following determine the clock cycles ($\mathcal{T}$) and average energy ($\mathcal{E}$) for the unit operation     
\begin{subequations} \label{Eq:power}
\begin{align}
\begin{split}
\mathcal{T} = W_P\times(1+2A_P)  
\end{split}\\
\begin{split}
\mathcal{E} = {} & W_P\times\Big(MC_{PL}V_{PCH}^2 \\
& + \sum_{i=0}^{A_P-1} E_{C} + E_{SAR} + 2^{i}C_{PL}V_{PCH}^2 \Big)
\end{split}
\end{align}
\end{subequations}
Here, $C_{PL}$ is the product line (PL) capacitance. $V_{PCH}$ is the precharge voltage for processing. $E_{C}$ and $E_{SAR}$ are the average energy of unit operation for the comparator and SA register logic. $M$ is the number of columns in each half of $\mu$Array. Since MAV operation requires all PL to charge, the estimated dynamic energy is $MC_{PL}V_{PCH}^2$ per weight bit plane. Within a bit plane processing, SA-ADC operates comparator and SAR logic $A_P$ times. From our simulations in 45 nm CMOS, Figure \ref{fig:surface}(d) shows various parameters in the equation. For $C_{PL}$, we add a 20\% overhead from the transistor capacitance to account for the interconnect parasitic.  

From the surface plots, $W_P$ and $A_P$ have different sensitivity to latency and energy demand on computing in our scheme. Consider two iso-accuracy cases: Case-A ($W_P$=8-bit, $A_P$=2-bit) and Case-B ($W_P$=4-bit, $A_P$=5-bit) in the surface plot. For both cases, the prediction accuracy is $\sim$95\%, but the cases are diametrically opposite in weight and ADC precision -- Case-A has the maximum W$_P$ and Case-B has the maximum A$_P$. For the iso-accuracy cases, Case-A has $\sim$10\% lower latency than Case-B while requires $\sim$30\% more energy. Case-B has lower energy due to fewer MAV steps, thereby, fewer MAV and ADC cycles. Case-A has lower latency since each MAV cycle is much shorter than in Case-B. Interestingly, an optimal $W_P$ and $A_P$ can be dynamically reconfigured in our design. $W_P$ can be lowered by skipping low significance weight bit-planes. $A_P$ can be lowered by limiting the successive approximation cycles of SA-ADC to a desired precision.  

\subsection{Impact of process variability and on-chip calibration}
\begin{figure}[t]
	\centering
    \includegraphics[width=\linewidth]{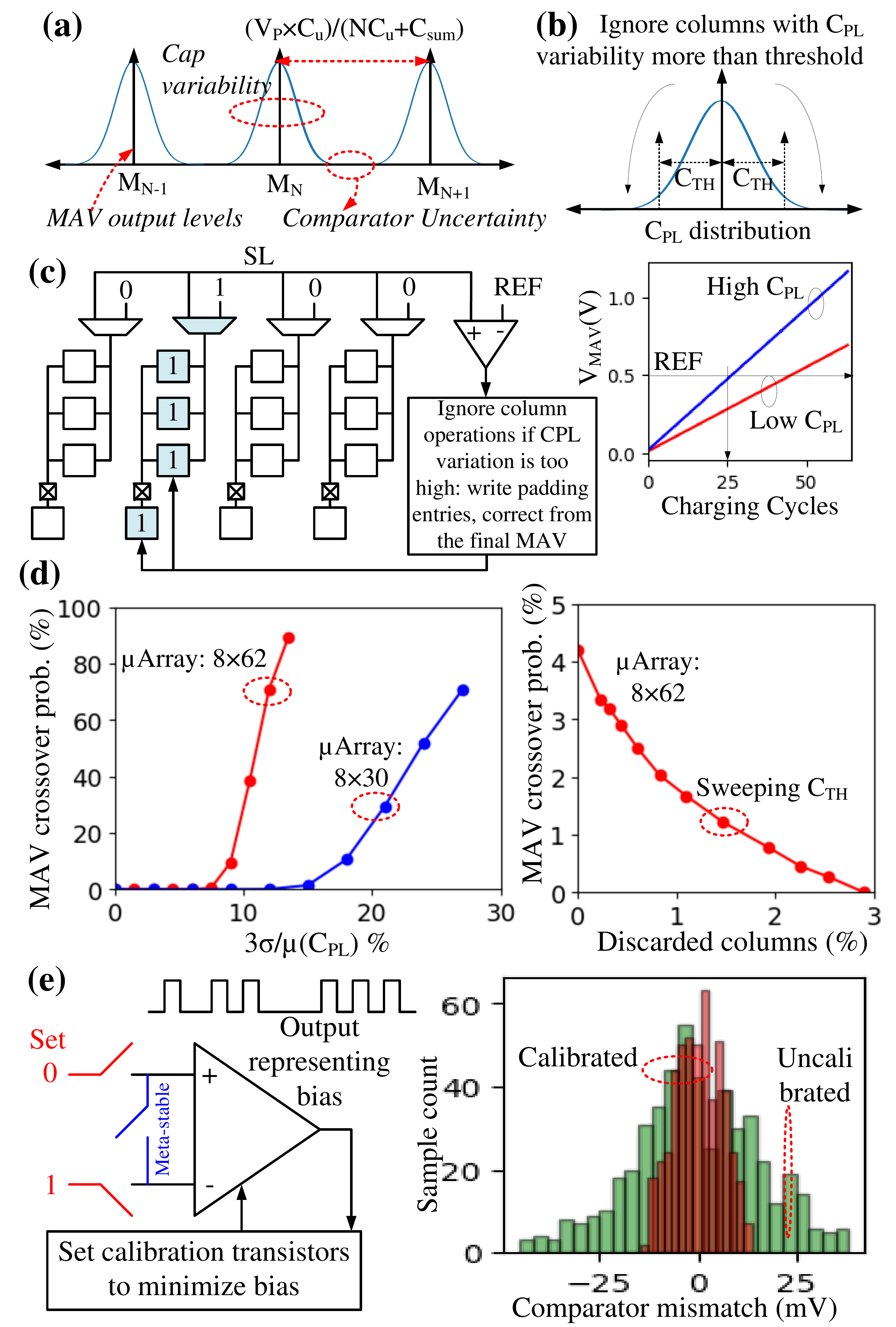}
	\caption{(a) MAV output levels vary due to process variability in PL capacitors. (b) $\mu$Array columns with extremely varying PL capacitor are discarded by padding them with memory and column entries that doesn't contribute to the MAV numerator. (c) An on-chip scheme to estimate PL columns with extremely varying capacitance. (d) MAV crossover probability at varying PL capacitor mismatch and $\mu$Array sizes. Mitigating MAV crossover probability by discarding columns with high PL capacitor variability. (e) Estimating comparator's variability by forcing it to metastable point and calibrating tail currents to mitigate process variability.}
	\label{fig:var}
\end{figure}
In Figure \ref{fig:var}(a), due to process variability among PL capacitors, MAV output levels will follow a Gaussian distribution. The distribution of MAV output levels arises both due to variability in PL capacitors as well as many combinations to obtain a MAV level. If MAV output levels crossover, the weight-input product from $\mu$Arrays can be erroneous. In our scheme, the accuracy of MAVs is mainly affected by the PL capacitor's mismatch. The effect of global variability among PL capacitors cancels out by bi-partitioning a $\mu$Array -- generating MAVs in one half and reference voltages in the other half -- so that the global variability of PL capacitors becomes common mode. Considering a Gaussian distribution of MAV output levels, Figure \ref{fig:var}(d) shows the probability of MAV crossover (P$_F$) in a $\mu$Array at varying capacitor mismatch and $\mu$array size. P$_F$ increases with higher PL capacitor variability as well as with the increasing number of columns in a $\mu$Array. Therefore, the maximum number of columns in a $\mu$Array (i.e., its parallelism) is constrained. 

In Figure \ref{fig:var}(c), we discuss an on-chip scheme to \textit{self-determine} the usable column width of a $\mu$Array based on its process variability. In the figure, the strength of a PL capacitor is measured on-chip by repeatedly charging the sum-line through it and counting the number of cycles to cross a set threshold. A smaller PL capacitor will require more charging cycles to cross the threshold. Most extreme PL capacitors are identified. If their process variability is more than an acceptable margin, these columns are not used [Figure \ref{fig:var}(b)]. In our scheme, we avoid adding a switch to disconnect such columns, since it will considerably increase the area overhead of the on-chip calibration scheme. Note that the column disconnect switch and a memory cell to store the switch enable needs to be implemented for each column of $\mu$Array. Instead, we lessen the effect of columns with extreme C$_{PL}$ variation by writing one to all SRAM cells in the column and by applying the CL input signal to be one. Therefore, the column with extremely varying C$_{PL}$ always discharges and only contributes to the charge averaging step. The sensitivity of extremely varying C$_{PL}$ to MAV is thereby low since it only contributes to the denominator of MAV, where its effect averages out against other columns in $\mu$Array. Based on this scheme, the right of Figure \ref{fig:var}(d) shows the MAV cross-over probability for 8$\times$62 $\mu$Arrays considering $\pm$12\% mismatch among PL capacitors and at varying C$_{TH}$ levels [Figure \ref{fig:var}(b)]. By discarding only about 3\% of columns, MAV cross-over probability can be sufficiently suppressed.     

Similarly, process variability in the comparator constraints the minimum precharge voltage and the maximum number of columns in a $\mu$Array. In Figure \ref{fig:var}(e), we use an on-chip calibration scheme to mitigate the comparator's process variability. The scheme selects N- and P-type counterparts of the comparator in turn. The comparator is first set to a known initial condition and then forced to a metastable point by shorting both inputs. By repeatedly resetting and setting the comparator, its bias can be estimated from the output bit sequence. An unbiased comparator should have an equal probability of 0/1 under thermal noise. The tail currents in the left and right half of the comparator can be adjusted to minimize the comparator's bias. Calibrating transistors for the comparator are shown in Figure 4(a). A counter monitors the comparator's output and adds calibration transistors to the left or right half to minimize bias in the comparator. In the right of Figure \ref{fig:var}(e), using a 2-bit calibration, the comparator's mismatch can be reduced to $\pm$12 mV from the initial $\pm$45 mV. 

\subsection{Comparisons to current art}
Table II compares our results against the state-of-art on different data sets. For various datasets, our network configuration was discussed in Sec. II. We achieve better prediction accuracy than the competitive approaches by processing with multibit precision weights and inputs, whereas many prior approaches were limited to binarized weights. Our 8$\times$62 SRAM macro, which requires a 5-bit ADC, achieves $\sim$105 tera operations per second per Watt (TOPS/W) with 8-bit input/weight processing at 45 nm CMOS. Our 8$\times$30 SRAM macro, which requires a 4-bit ADC, achieves $\sim$84 TOPS/W. SRAM macros that require lower ADC precision are more tolerant of process variability, however, have lower TOPS/W as well. Our implementation achieves higher energy efficiency and performance by co-adapting DNN operators with SRAM's implementation and operational constraints. Notably, our approach is DAC-free. Our bitplane-wise processing also lowers the necessary ADC's precision. Area overheads in our ADC minimize by exploiting bit line capacitances. Due to these cross-cutting design transformations from computing primitives to micro-architecture, our 8$\times$62 $\mu$Array-based platform achieves 15$\times$ higher macro-level energy efficiency than \cite{1st} even though the latter is in 28 nm CMOS. Compared to 65nm design in \cite{2nd}, it achieves 35$\times$ better TOPs/W. Although our efficiency is lower than \cite{5th}, note that the latter is only for 1-bit inputs and weights. The accuracy of 1-bit processing in \cite{5th} suffers on more complex processing tasks, such as for CIFAR-10 in the table. 

\begin{figure}[t]
	\centering
    \includegraphics[width=0.9\linewidth]{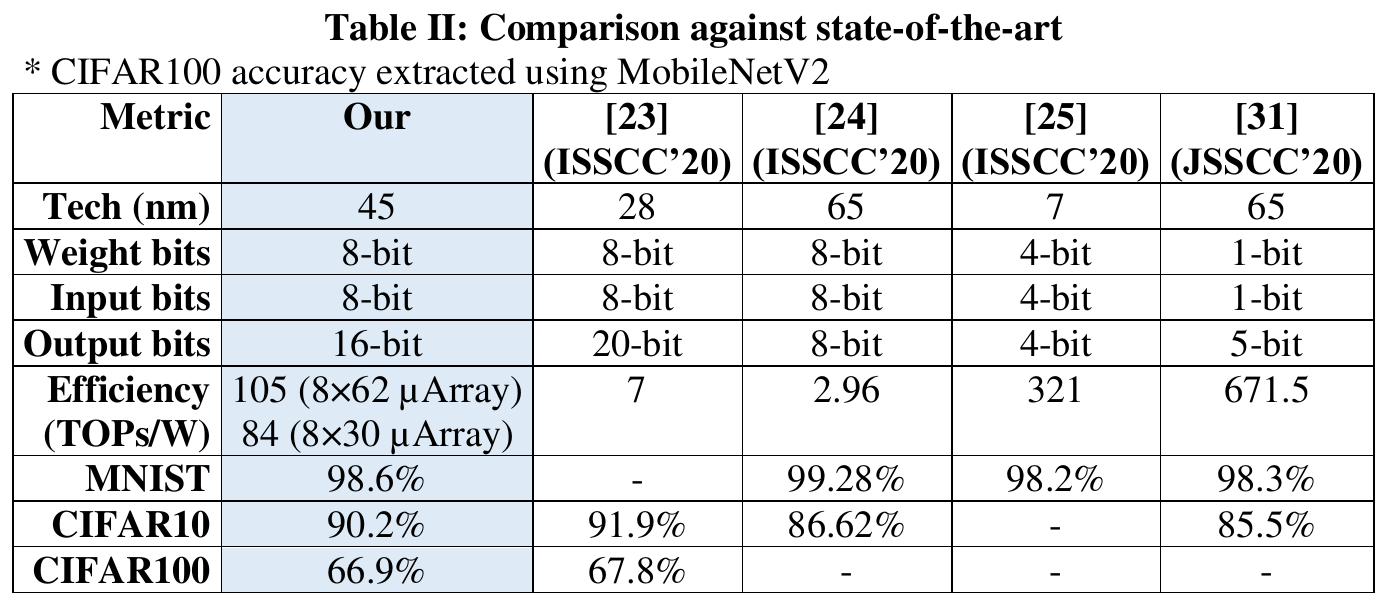}
\end{figure}

\section{Synergistic Integration of Digital and Compute-in-Memory Processing for DNN}
Compute-in-memory offers immense energy efficiency benefits over digital by eliminating weight movements. Mixed-signal processing of compute-in-memory also obviates processing overheads for adders by exploiting physics (Kirchoff's law) to sum the operands over a wire. Note that additions are a significant portion of the total workload in a digital DNN inference. However, compute-in-memory is also inherently limited to only weight stationary processing. The advantages of stationary weight processing reduce if the filter has fewer channels or if the input has smaller dimensions. Compute-in-memory is also more area expensive compared to digital processing, which can leverage denser memory modules such as DRAM. On the other hand, the memory cells in compute-in-memory are larger to support both storage and computations within the same physical structure. Additionally, multibit precision DNN inference is complex using compute-in-memory. Therefore, many prior works \cite{courbariaux2016binarized,5th} utilize binary-weighted neural networks, which, however, constraints the learning space and reduces the prediction accuracy. Deep in-memory architecture (DIMA) in \cite{Kang} considers multibit precision in-memory inference; however, the implementation suffers from an exponential reduction in the throughput with increasing precision. Meanwhile, in this work, we overcame the critical challenge using a novel co-design approach by adapting the DNN operator to in-memory processing constraints. In our multiplication-free compute-in-memory framework, the parametric learning space expands, yet the implementation complexities are equivalent to a binarized neural network. Even so, the accuracy of multiplication-free operators is somewhat lower than the typical deep learning operator due to the non-differentiability of gradients. 

\begin{figure}[t]
\centering
\subfloat[]{\includegraphics[width=0.85\linewidth]{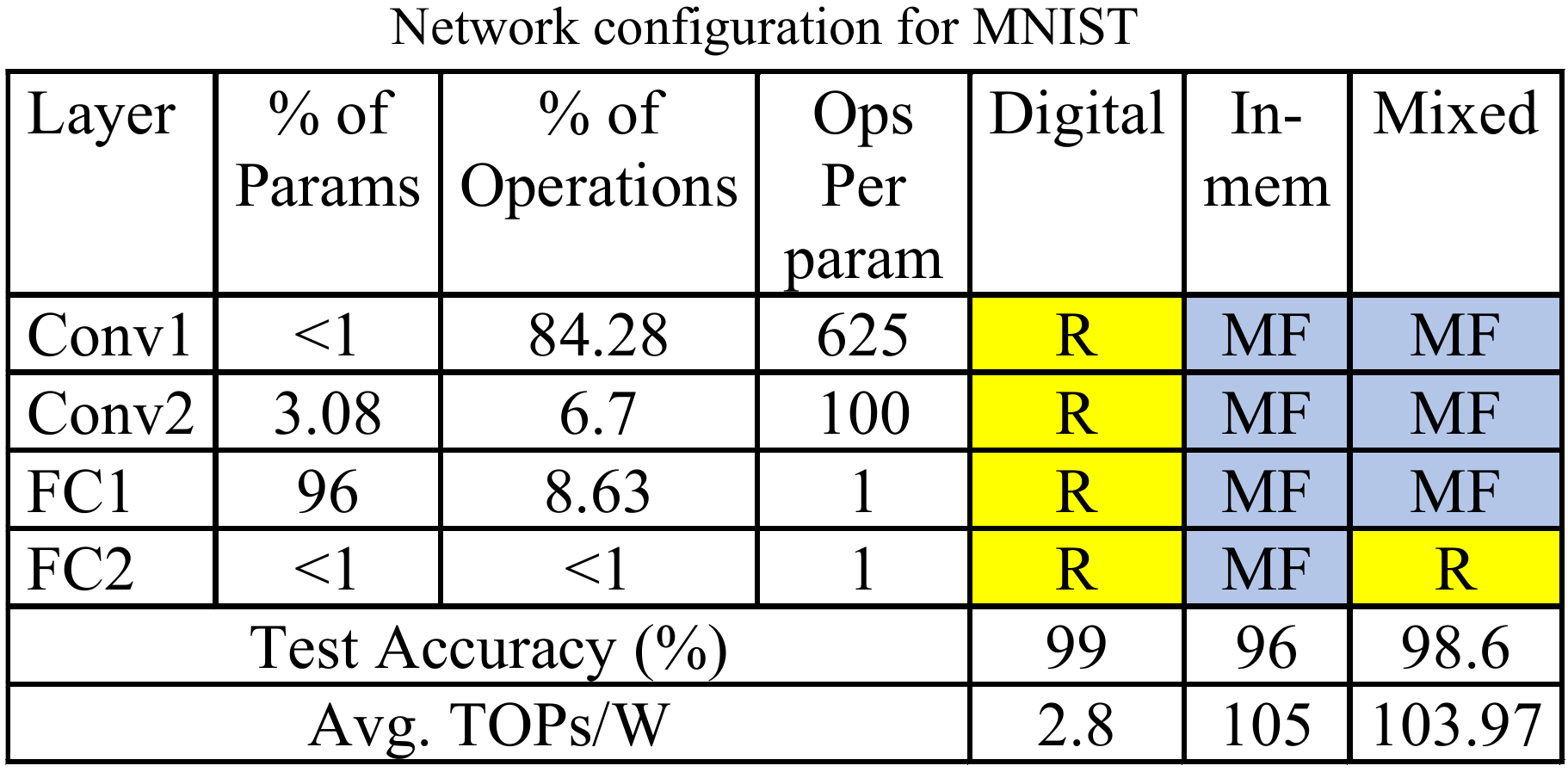}}\\
\subfloat[]{\includegraphics[width=0.85\linewidth]{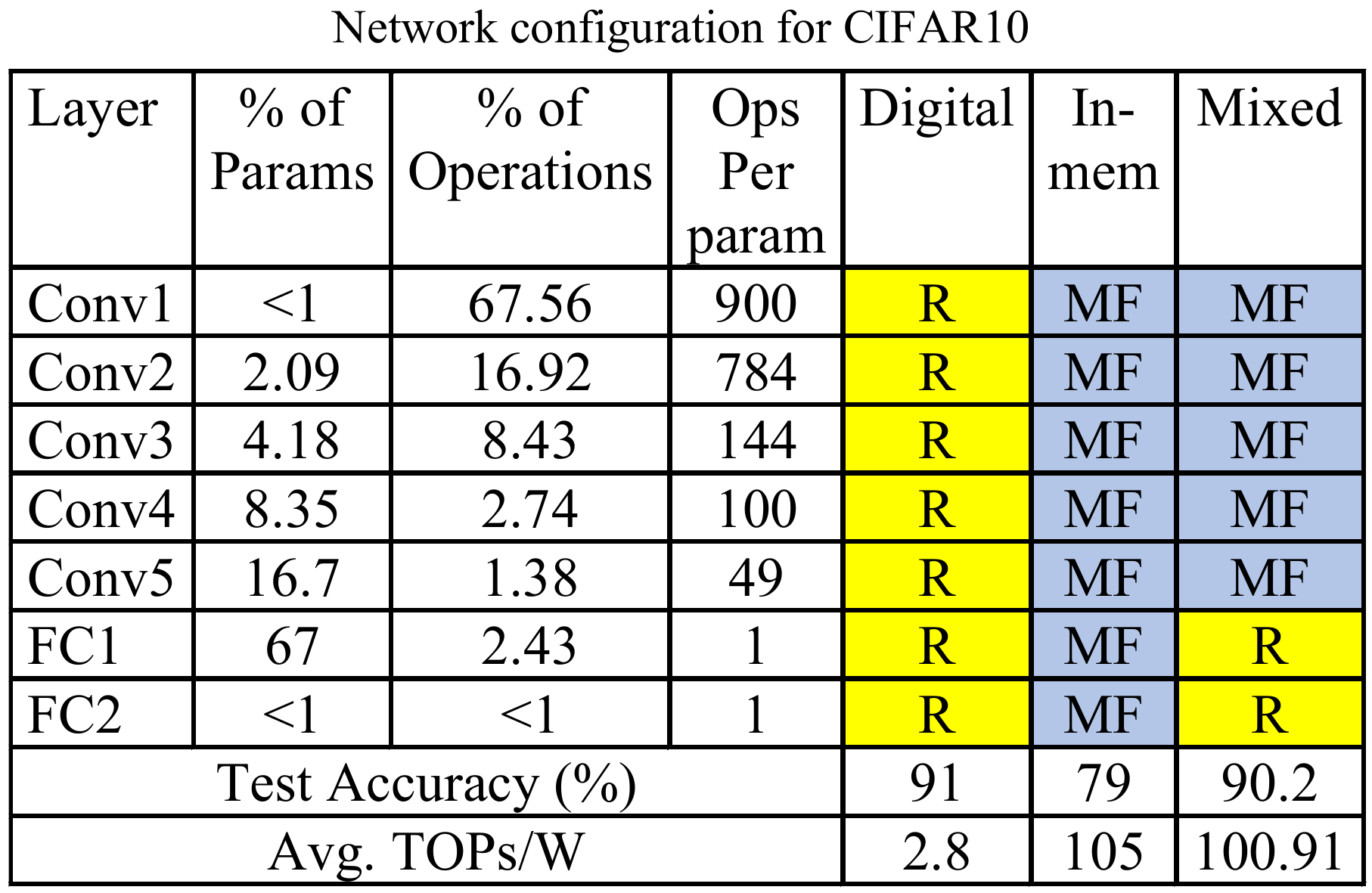}}\\
\subfloat[]{\includegraphics[width=0.85\linewidth]{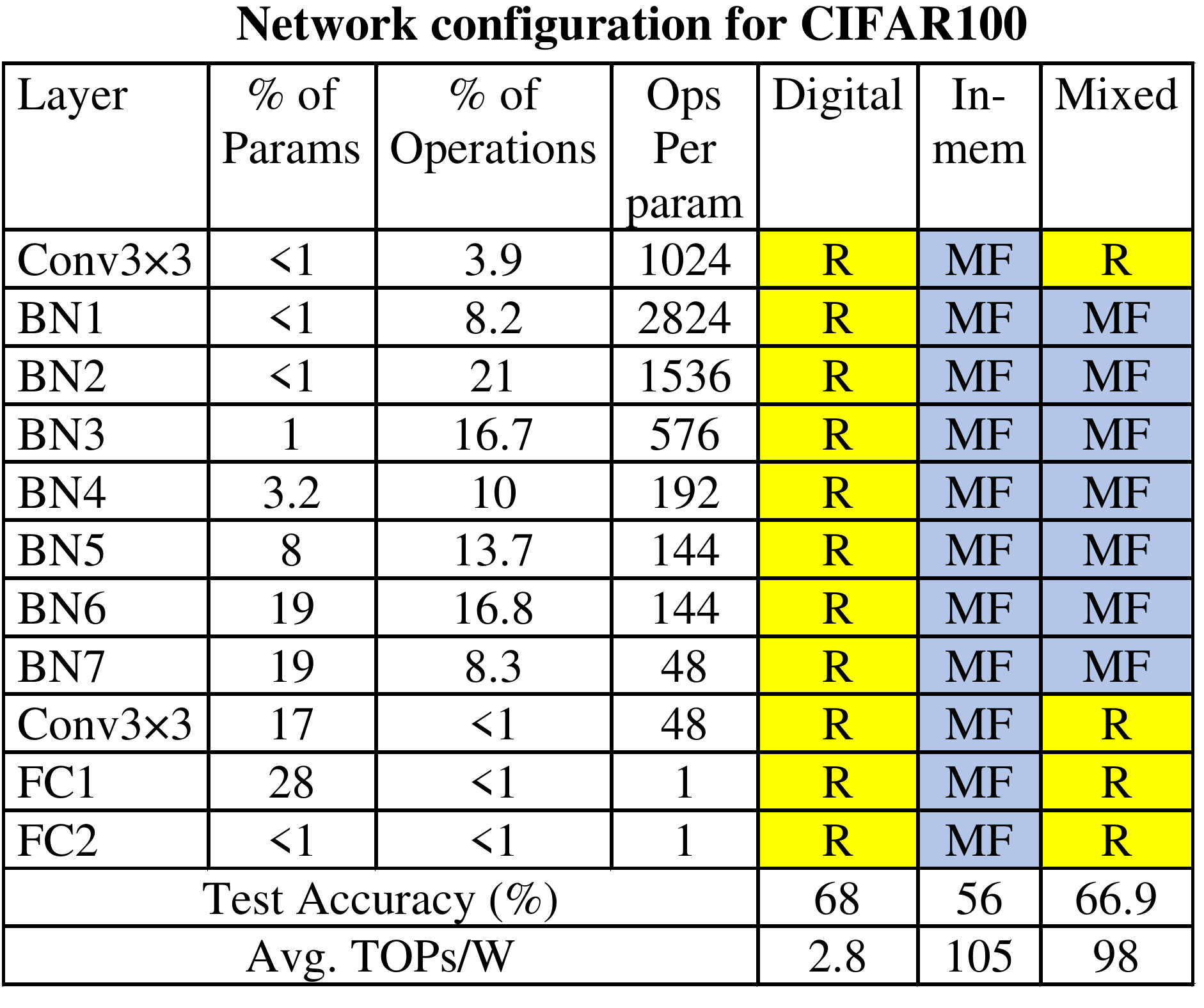}}

\caption{Synergistic mixing of compute-in-memory and digital processing for (a) MNIST, (b) CIFAR10, and (c) CIFAR100. }
\label{fig:config}
 \end{figure}

Considering the above trade-offs, we find that the key to balance scalability with energy efficiency in DNN inference is through a synergistic integration of compute-in-memory with digital processing. In Figure \ref{fig:config}, we discuss insights towards this. The figure shows a layer-wise distribution of weights and operations for the networks used for MNIST, CIFAR10, and CIFAR100. The network configurations were discussed earlier in Sec. II. Note that as the processing propagates through the networks, weights per layer increase, but the number of operations per weight reduces. This is, in fact, typical to any DNN due to shrinking input feature map dimensions, which reduces the weight reuse opportunities. Since the starting layers have much fewer parameters but much higher weight reuse, they are quite suited for compute-in-memory. The latter layers require many more parameters but have low weight reuse. Therefore, digital processing can minimize the excessive storage overheads of these layers with denser storage.

Using this strategy, the figure also shows a \textit{mixed} mapping configuration that layer-wise combines compute-in-memory and digital processing. For example, in the mixed implementation of MobileNetV2, feature extraction layers with high weight reuse are mapped in compute-in-memory using an 8-bit multiplication-free operator. Regression layers and others with low weight reuse are mapped in digital using the typical operator. Remarkably, based on the synergistic mapping strategy, compute-in-memory only stores about a third of the total weights; yet, performs more than 85\% of the total operations. Therefore, the synergistic mapping can optimally translate compute-in-memory’s energy-efficiency advantages to the overall system-level efficiency, and yet, limits its area overheads. The synergistic mapping also improves the prediction accuracy, since only critical layers are implemented with the energy-expensive typical operator while the remaining most of the network is operated with multiplication-free operators. The figure also shows similar mapping configurations for MNIST and CIFAR10 prediction networks. In the figure, we also project average macro-level energy efficiency in TOPs/W. For digital processing, we use 2.8 TOPs/W from \cite{Tops}. For multiplication-free compute-in-memory processing, we use 105 TOPs/W from Table II. We, however, note that the macro-level energy efficiency doesn't necessarily translate to system-level energy efficiency where overheads due to routing and control flow must also be accounted for. We plan to consider this characterization in future work. 

\section{Conclusion}
We presented a compute-in-SRAM macro based on a multiplication-free learning operator. The macro comprises low area/power overhead $\mu$Arrays and $\mu$Channels. Operations in the macro are DAC-free. $\mu$Arrays exploit bit line parasitic for low overhead memory-immersed data conversion. We characterized the accuracy of our scheme on MNIST, CIFAR10, and CIFAR100 data sets.  Other researchers successfully used the multiplication-free operator based structures in various other datasets\cite{pan,ergen}. On an equivalent network configuration, our framework has 1.8$\times$ lower error on MNIST and 1.5$\times$ lower error on CIFAR10 compared to the binarized neural network. At 8-bit precision, our 8$\times$62 compute-in-SRAM $\mu$Array achieves $\sim$105 TOPS/W, which is significantly better than the current compute-in-SRAM designs at matching precision. Our platform also offers several runtime control knobs to dynamically trade-off accuracy, energy, and latency. For example, weight precision can be dynamically modulated to reduce prediction latency, and ADC's precision can be controlled to reduce energy. Additionally, for deeper neural networks, we have discussed mapping configurations where high weight reuse layers can be implemented in our compute-in-SRAM framework, and parameter-intensive layers (such as fully-connected) can be implemented through digital accelerators. Our synergistic mapping strategy combining both multiplication-free and typical operator is promising to achieve both high energy efficiency and area efficiency in operating deeper neural networks.

\balance
%\nocite{*} % to test all bib entrys
\bibliographystyle{IEEEtran}
\bibliography{refs}
\end{document}